# Single-shot fourth-order auto-correlator

**Peng Wang[1, a] Xiong Shen[1, a] Jun Liu[1, 2, 3, *] and Ruxin Li[1, 2, 3, *]**

[1]State Key Laboratory of High Field Laser Physics, Shanghai Institute of Optics and Fine Mechanics, Chinese Academy of Sciences, Shanghai 201800, China

[2]University Center of Materials Science and Optoelectronics Engineering, University of Chinese Academy of Sciences, Beijing 100049, China

[3]CAS Center for Excellence in Ultra-intense Laser Science (CEULS), Shanghai Institute of Optics and Fine Mechanics, Chinese Academy of Sciences, Shanghai 201800, China

[a] These authors contributed equally to the work.

**Abstract:** The temporal contrast is one of the most important parameters of an ultra-high intense laser pulse. Third-order auto-correlator or cross-correlator have been widely used to characterize the temporal contrast of an ultra-intense laser pulse in the past decades. Here, a novel and simple single-shot fourth-order auto-correlator to characterize the temporal contrast with higher time resolution and better pulse contrast fidelity in comparison to third-order correlators is proposed. The single-shot fourth-order autocorrelation consists of a frequency degenerate four-wave mixing process and a sum-frequency mixing process. The proof-of-principle experiments show that a dynamic range of approximately $10^{11}$ compared with the noise level, a time resolution of approximately 160 fs, and a time window of 65 ps can be successfully obtained using the novel single-shot fourth-order auto-correlator, which is the highest dynamic range with simultaneous high time resolution for single-shot temporal contrast measurement so far. Furthermore, the temporal contrast of laser pulse from a PW laser system is successfully measured in single-shot with a dynamic range of about $2\times10^{10}$ and simultaneous a time resolution of 160 fs.

*Corresponding author: jliu@siom.ac.cn; ruxinli@mail.siom.ac.cn

## 1. Introduction

Many petawatt (PW) laser systems with the benefit of chirped pulse amplification (CPA) [1] and optical parametric chirped pulse amplification (OPCPA) [2] techniques have been demonstrated in the past decade [3]. State-of-the-art high-power laser systems can produce intense pulses with peak powers of 10 PW [4]. In the future, laser systems may reach the 100-PW level [5] with focal intensities of $10^{21}$–$10^{24}$ W/cm$^2$ [6]. Ultra-high intense laser pulses have been applied in the field of laser-matter interaction [7], such as proton or electron acceleration in thin solid targets [8-10] and electron generation in fast-ignition inertial confinement fusion [11, 12]. For these applications, the temporal contrast (TC) of the ultra-intense laser pulse is one of the most important parameters. Because the ionization threshold intensity of most solid targets is in the $10^{11}$ W/cm$^2$ range, the TC of ultra-intense laser pulses should be higher than $10^{10-13}$ to avoid the damage of target by the high intensity background or prepulses. Recently, compressed ultra-high intense laser pulses with $10^{12}$ TC have been developed [13]. Because most PW laser systems run at low repetition rates or even single-shot, then single-shot characterization of the TC of an ultra-intense laser pulse is crucial.

However, although many techniques have been developed to improve the TC of ultra-intense laser pulses [14-16], only a few methods have been proposed for single-shot characterization of the TC. The first and most frequently used method for the measurement of the TC of single-shot pulses [17] is third-order auto-correlator (TOAC). A time window of approximately 200 ps and $10^6$ dynamic range was obtained by using a pulse replicator in the TOAC [18]. By using the optical parametric amplification (OPA) process to generate the sampling pulse, a dynamic range as high as $10^{10}$ and a 50-ps time window with a 700-fs time resolution was achieved by the Qian group based on a fiber-array-based detection system [19]. With typical TOAC and third-order cross-correlator (TOCC) processes, it is hard to achieve single-shot pulses with simultaneous high dynamic range, precise time resolution, high pulse contrast fidelity, and wide time window, which are the most important parameters for TC characterization. Recently, the SRSI-ETE method was demonstrated, which has a time resolution as high as 20 fs [20-22]. However, the dynamic range is limited by the signal to noise ratio of the detector and, at present, the highest achievable value is $10^8$. Together with the temporal contrast reduction method, $10^9$ dynamic range can be obtained by using SRSI-ETE recently [23].

In this paper, we propose a novel single-shot measurement method named fourth-order autocorrelation for the single-shot TC characterization of ultra-intense laser pulses. In this method, frequency-degenerate third-order nonlinear processes, such as cross-polarization wave generation (XPW) [16,24], self-diffraction (SD) [25, 26], and transient grating (TG) [27], have the potential to generate cleaner signal pulses, which can then be used as the cleaning reference (or sampling) pulses for the next sum-frequency mixing (SFM) process together with the test pulse. In the proof-of-principle experiment, the SD process is used to generate the cleaning sampling pulse in the novel single-shot fourth-order auto-correlator (FOAC). A single-shot measurement with $10^{11}$ dynamic range, 65-ps time window, and 160-fs time resolution has achieved.

Compared with previous single-shot TOAC or TOCC methods, the single-shot FOAC has the following advantages. First, cleaner sampling pulses are generated by the third-order nonlinear processes and as a result, higher fidelity measurement is achieved. Second, the sameness of the sampling pulse and test pulse central wavelengths results in a high time resolution of the single-shot measurement owing to the neglecting influence of the group velocity mismatch (GVM) during the SFM process. Third, hundreds of μJ sampling pulses are generated by a simple SD process, and a $10^{11}$ dynamic range is achieved. Furthermore, two improvements are made in the proof-of-principle experiment to reduce the noise and extend the time window. First, the Beta-Barium Borate (BBO) crystal for SFM is cut with a large angle to reduce the noise from the second-harmonic generation (SHG) signals of both the sampling pulse and test pulse. Second, the aperture of the BBO crystal for SFM is extended to a width of 21 mm, which can support a 65 ps time window in single-shot measurement.

## 2. Principle of fourth-order autocorrelation

To further explain the principle of fourth-order autocorrelation, we will compare it with second- and third-order correlation, which are used in the characterization of pulse duration and TC, respectively. In the second-order autocorrelation, the ultra-short pulse to be measured, which is shown in Figure 1(a), is used as its own sampling pulse by splitting the pulse into two replicas I(t) and I(t-τ). As a result, the corresponding autocorrelation signal is given by

$$S_2(\tau) = \int_{-\infty}^{\infty} I(t) I(t-\tau) dt \qquad (1)$$

where $\tau$ is the relative time delay. Then, second-order autocorrelation signal is always symmetric in time, and asymmetric pulses cannot be recognized by second-order autocorrelation, as shown in Figure 1 (b).

To measure an asymmetric pulse shape and the TC of an ultra-intense laser pulse, third-order auto-correlation was proposed in 1993 [17]. First, the sampling pulse is obtained by using the typical second harmonic generation of the test pulse. Then, it is followed by a frequency nondegenerate SFM process, which is also a second-order nonlinear process. The corresponding signal can be expressed as

$$S_3(\tau) = \int_{-\infty}^{\infty} I^2(t) I(t-\tau) dt \qquad (2)$$

where $\tau$ is the relative time delay. Obviously, third-order autocorrelation signal $S_3(\tau)$ is asymmetric, and the prepulses and postpulses can be distinguished clearly, as shown in Figure 1(c).

At present, commercial TOACs based on delay-scanning optical setups can guarantee a $10^{10}$ dynamic range capability. With the usage of noise filtering, high sensitivity detector, and attenuator, dynamic ranges as high as $10^{13}$ can be achieved using delay-scanning TOAC [28]. Because the wavelength of the third-order autocorrelation signal is different from those of the SHG signals of the two incident pulses, it can suppress the scattering noise from the two incident pulses by using a spectral filter, which leads to a high dynamic range. However, two problems are still faced. Firstly, the time resolutions of the delay-scanning TOACs are limited to about 250 fs according to previous works [19,20], mainly due to the GVM induced by the large central wavelength difference between the sampling pulse and the test pulse and the thickness of the nonlinear crystal. Secondly, strong postpulses will still induce ghost prepulses in the third-order correlation signal, as shown in Figure 1(c). As the dynamic range of TC becomes higher and higher, the ghost prepulses become detectable, which will affect the fidelity of the measurement.

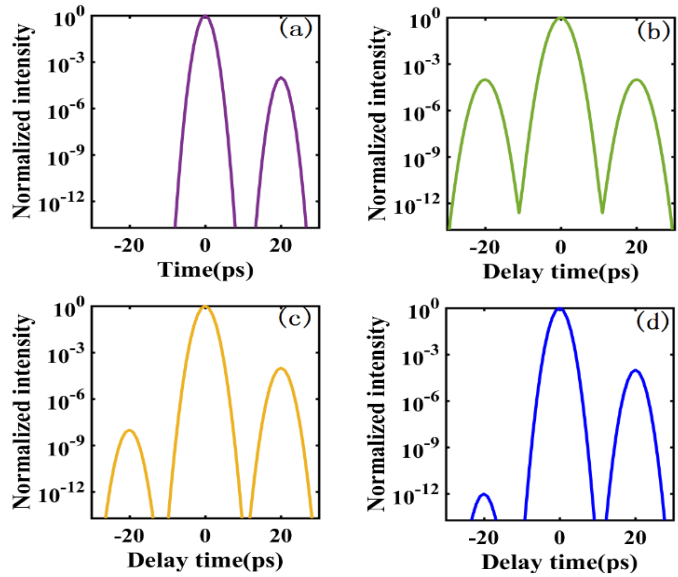

Figure 1. (a) The testing pulse with a postpulse. (b) The second-order autocorrelation signal. (c) The third-order autocorrelation signal. (d) The fourth-order autocorrelation signal.

To solve these two problems, fourth-order autocorrelation is proposed here. In the fourth-order autocorrelation, the sampling pulse is generated by a frequency degenerate four-wave mixing processes, such as SD, XPW, and TG, which are third-order nonlinear processes. Then, the fourth-order autocorrelation signal is obtained by the following SFM process. As a result, a third-order nonlinear process and a second-order nonlinear process make up the fourth-order autocorrelation, which can be expressed as

$$S_4(\tau) = \int_{-\infty}^{\infty} I^3(t) I(t - \tau) dt \qquad (3)$$

where $\tau$ is the relative time delay. If the TC of the input pulse is R, the third-order sampling pulse will have a TC of $R^3$. As a result, the TC measurement has a higher fidelity for FOAC than for TOAC. Owing to the high TC of the sampling pulse obtained by third-order nonlinear process, the ghost prepulse induced by the strong postpulse in third-order autocorrelation is greatly diminished, as shown in Figure 1(d). Moreover, similar to second-order autocorrelation, the final SFM process of fourth-order autocorrelation is also frequency degenerate, which leads to a high time resolution of TC measurement owing to a negligible GVM limitation.

3. **Single-shot FOAC**

As single-shot characterization of the TC is necessary for most PW laser systems, which are run at low repetition rates or even single-shot. Single-shot FOAC will be demonstrated here. Different from delay-scanning setup, both single-shot second-order auto-correlator and single-shot TOAC encode time into space by using a special beam-crossing geometrical arrangement in the final SFM process [19, 29]. In single-shot TOAC or TOCC, the two crossing beams have different central wavelengths, which can suppress the SHG scattering noise of the two incident pulses by using dichroic short-pass filers. However, the large difference in wavelength will lead to a relatively large GVM in SFM process, which will seriously affect the time resolution of single-shot auto-correlator. The time resolution can be expressed as $l \times \left(\frac{1}{v_s \cos\theta} - \frac{1}{v_p \cos\alpha}\right)$, where $l$ is the thickness of the nonlinear crystal, $v_s$ and $v_p$ are group velocities of the sampling pulse and the test pulse, respectively, and θ and α are the crossing angles of the test pulse and sampling pulse, respectively. As a result, the thickness of the nonlinear crystal for SFM should be chosen to balance the high time resolution and intense autocorrelation signal, which will affect the intensity improvement of the autocorrelation signal.

Different from single-shot TOAC or TOCC, in the proposed single-shot FOAC the sampling pulse has the same wavelength as the test pulse. Therefore, the influence of GVM to the time resolution is removed as shown in the Figure 2. Then a thicker nonlinear crystal can be used to obtain relatively high correlation signal, which will result in a measurement with high dynamic range. Previous works have shown that the frequency degenerate four-wave mixing processes XPW [16,24], SD [25, 26], and TG [27] satisfy this requirement very well. These three processes achieve a cleaning pulse with the same wavelength. Among them, the XPW [30] and SD processes [31] have been used to obtain seed pulses with high TC in ultra-high intense laser systems. SD has the ability to improve the TC by seven orders of magnitude in single glass plates [25]. This is the highest TC enhancement achieved with a single-stage process. Furthermore, more than 500 μJ first-order SD signal can be obtained in a glass plate [25]. As a result, only the SD process is explored in the following proof-of-principle experiments.

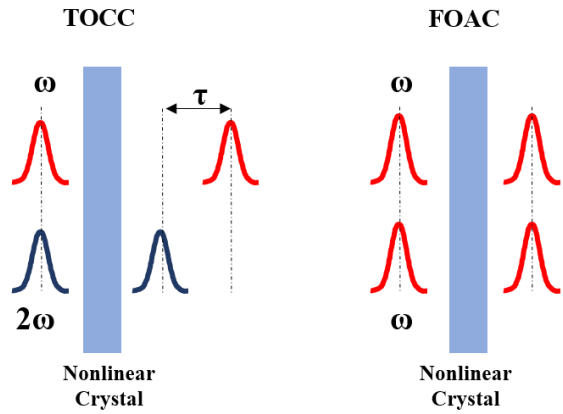

Figure 2. The GVM in TOCC and FOAC.

## 4. Optical setup and experiment

The optical setup of FOAC using SD process is shown in Figure 3. The input beam is split by a beam splitter with a 30:70 reflection/transmission (R:T) ratio. The transmitted beam is used for the clean sampling pulse generation based on the SD process, while the reflected beam is used as the test pulse. The transmitted beam is further split by a beam splitter with a 50:50 R:T ratio. After the splitter, the two beams for SD process are focused by two cylindrical lenses and overlapped in a glass plate with an overlapping area of approximately $1 \times 13$ mm. Bright sidebands emerge when the two input laser are overlapped spatially and temporally. The first-order SD signal is picked out as the sampling pulse, which is then collimated by a concave cylindrical mirror. Both the high energy sampling pulse and the test pulse are expanded using two beam expanders. Then, they are focused into a wedge-designed nonlinear BBO crystal by two 200-mm focal-length concave cylindrical reflective mirrors. The 21-mm-wide BBO is cut with the angle of 80° and the two input beams are intersected with a crossing-angle of 63° in air. Finally, the FOAC signal is produced by SFM process in the BBO crystal, which is imaged and then recorded by using a 16-bit sCMOS camera through a 4f lens pair. The sCMOS camera has 2048 × 2048 pixels with 11 × 11 μm pixels and nearly single photon detect sensitivity (Tucsen Photonics, Dhyana 95). To facilitate the analysis of the correlation signal, a strip-shaped density filter, with an approximately 4 orders of magnitude attenuation of the main pulse, and a coated wedge which introduces an attenuated reference replica signal below the original correlation signal, are placed behind the BBO crystal. The strip-shaped density filter with a metallic coating has a fixed attenuation ratio about $10^4$ for the SFM signal around 400 nm, which has been calibrated before the experiment. The coated wedge would introduce a reference replica signal with about 70 times attenuation own to the back-and-front reflective on the surface of the wedge. Because of the diffraction, the edge effect would influence the signal near the edge of the strip-shaped density filter. However, the density filter is placed right behind the BBO crystal and the correlation signal is imagined from the BBO crystal to the sCMOS sensor by a 4f imaging system. By carefully adjust the 4f imaging system, the influence of the diffraction effect would be weakened. A band-pass filter with a central wavelength of 400 nm is placed in front of the sCMOS camera to avoid noise from the fundamental incident pules. The whole setup is compact and economical as a single-shot measurement apparatus.

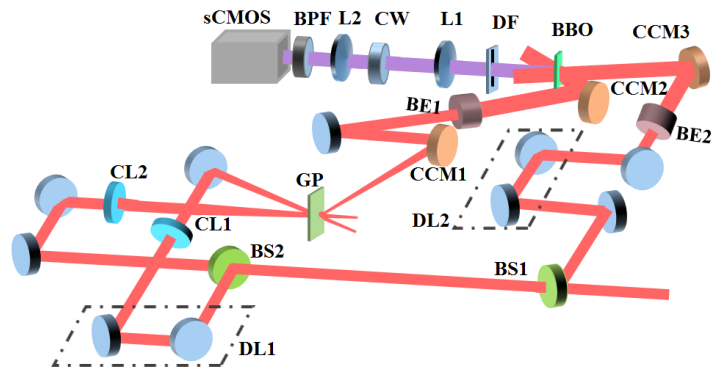

Figure 3. Schematic representation of the proof-of-principle experimental setup using FOAC for single-shot TC measurement. BS1-2 are two beam splitters. CCM1-3 are three concave cylindrical mirrors. DL1-2 are two time delay lines. GP is a glass plate for SD signal generation. BE1 and BE2 are two beam expanders. DF is a strip-shaped density filter. CW is a coated wedge. L1 and L2 are two lenses that form the 4f imaging lenses. BPF is a band-pass filter.

## 5. Experimental results and discussion

A femtosecond Ti: sapphire regenerative amplifier ((8mJ/1 kHz/40 fs) is used to test the novel fourth-order autocorrelation method and the single-shot FOAC device at first. A high autocorrelation signal means a high dynamic range of the single-shot measurement. Considering the low energy transfer efficiency of SD process, laser pulses of approximately 8 mJ with a diameter of about 10 mm are all guided into the experimental setup. After the first beam splitter, the 2-mJ laser pulse is used as the test pulse and the left 6-mJ is used for sampling pulse generation. Intense SD sidebands are generated when the two input beams intersect each other with a 1.3° angle in air and synchronously overlapped in both time and space. The pulse energy of the first-order SD signal is measured to be 150 μJ with an efficiency of 2.5%, which can support a high dynamic range single-shot TC measurement. The generated SD signal has a smoother spectrum and cleaner TC than those of the input pulses. More details about the generation of the sampling pulse based on the SD process can be found in our previous paper [25].

Then, the sampling and test pulses are focused onto a large wedge-designed BBO crystal. The BBO has a central thickness of 1.5 mm to increase the intensity of the correlation signal, and then to improve the measureable dynamic range at the expense of decreased phase matching bandwidth. By capturing the spectrum of the SFG signal, the bandwidth of the SFG process is estimated to be about 20 nm. If the test pulse has a broadband spectrum, the phase matching bandwidth may not broad enough to cover the whole spectrum in the SFG process with a thicker nonlinear crystal. Fortunately, the spectral bandwidths of the main pulse, the pre- or post- pules, and the ASE background or fluorescence noise usually have the same spectral bandwidth, the SFG signal intensities of them responses to the phase matching bandwidth narrowing during SFG process are equal. Then the generated SFG signal with a narrow bandwidth using a thicker nonlinear crystal would has negligible influence to the temporal contrast ratio. Because the time window is determined by the crossing angle and the diameter of the BBO, then the BBO is designed to have a large aperture of 21 mm and a cutting angle of 80°, which supports the phase matching angle of 32° in the crystal. Furthermore, in the experiment, the relatively large cutting angle reduces the noise from SHG signals of both sampling and test pulses.

By adjusting the time delay between the sampling and test pulses, the autocorrelation signal of the main pulse can be observed as a bright blue spot on a white paper. The bright blue spot moves along the horizontal direction as the time delay changes. This is because the overlapping zone between the sampling and main pulses of the test pulse changes continuously with the time delay. The intense correlation signal guarantees a high dynamic range of the measurement, which has to be preprocessed due to the limited dynamic range of the sCMOS camera. Several steps are involved in obtaining a high dynamic TC measurement. Firstly, the correlation signal from the main pulse of the test pulse is so intense that a strip-shaped density filter is added after the BBO crystal to attenuate the intensity of the main pulse directly. However, some strong prepulses or postpulses can still saturate sCMOS camera. In that case, a coated wedge is added, which will introduce a reference replica signal with about 70 times attenuation in the shifted area of the sCMOS sensor. As a result, the original authentic intensity of the saturated main pulse and strong satellite pulses can be retrieved with the reference replica. Finally, the autocorrelation signal passes through a bandpass filter and is sent into the sCMOS.

Figure 4. (a) shows the fourth-order autocorrelation signal obtained using the sCOMS camera when the exposure time is 1 ms. The intense signal is the original autocorrelation signal of which the strongest signal from the main pulse has already been attenuated by a strip-shaped density filter. The weaker signal line below without saturation is the attenuated replica signal introduced by the front-and-back reflection of the coated wedge, which can be used to obtain the original real intensity of the main pulse and the strong postpulses or prepulses.

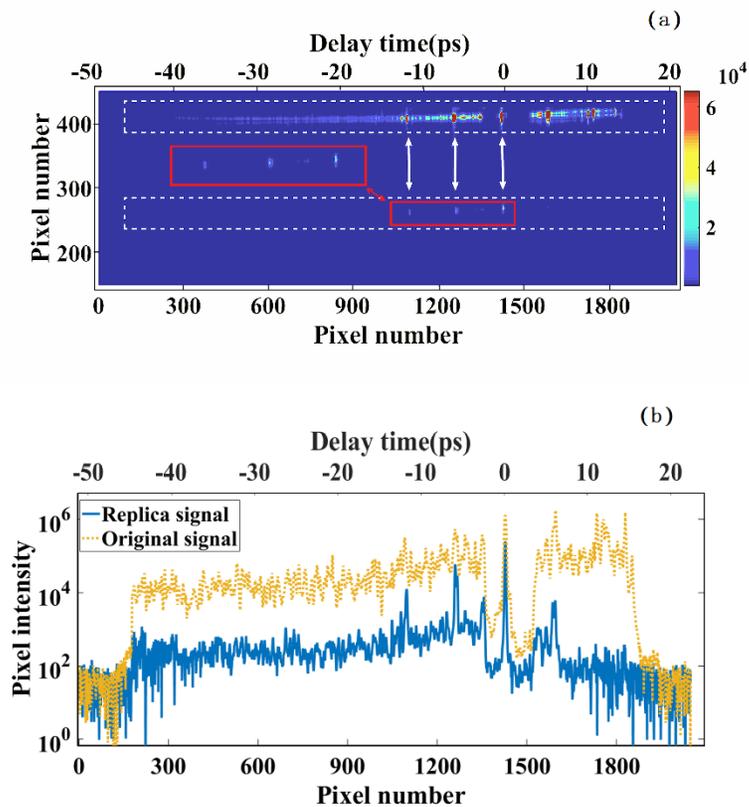

Figure 4. (a).The intensity distribution of the original signal and the replica on the sCMOS . The original signal and the replica are shown in the rectangle area marked with white dash lines. Part of the replica signal marked with red lines is zoomed to make it more visible as shown in the red rectangle area above the original one. (b) The intensity distribution of the original signal and the replica along the horizontal direction.

The bright correlation signal detected by the sCMOS camera covers an area of approximately $20 \times 1770$ pixels in the vertical and horizontal directions, respectively, where the time window is encoded into the horizontal direction. The time window can be calculated by $2 \times l \times p \times sin\frac{(\theta)}{c}$, where $l$ is the pixel size of the sCMOS sensor, $p$ is the pixel number relative to the center of the main pulse, $\theta$ is the crossing angle of the sampling pulse and the autocorrelation signal in air, and $c$ is the speed of light in air. As a result, the time window is approximately 65 ps in this experiment and each 11-μm pixel corresponds to approximately 37 fs. After summing up the intensity of the 20 pixels along the vertical direction and subtracting the background noise of the sCMOS, the intensity profile along the horizontal direction is obtained, as shown in Figure 4. (b) . Obviously, pixel numbers between 1372 and 1516 or delay times between -2 and 3.2 ps are attenuated by the strip-shaped density filter. In the vertical direction, all the 20 pixel for the main pulse are all saturated of the original SFG signal. However, for those pre-pulses or post-pulses in the original SFG signal, only a few pixels get saturated as shown in the Figure 4. (a), not all the 20 pixels are saturated. Therefore, even though the main pulse shows a saturation peak, the pre- and post- pulses do not have the same peak level, as shown in the Figure 4. (b). Same method is used to process the data of the replica signal in the Figure 4. (a), and the intensity profile of the weaker replica signal is obtained as shown in Figure 4.(b) (blue solid line). Thus, the intensity of the saturated correlation peaks can be retrieved correctly.

The real intensity of the main pulse is obtained by multiplying the attenuation ratio of the strip-shaped density filter and the front-and-back reflection of the coated wedge. Furthermore, two shots of FOAC with shifted times are measured and combined together to further extend the time window. The final TC profile measured by the novel FOAC is shown in Figure 5(dotted line). The TC of the kHz laser pulses has a dynamic range of $10^8$, and the time window with two shots is extended to approximately 106 ps with 48 and 58 ps in the front and back edges, respectively. The $10^{10}$ dynamic range at both ends of time window shows the background of sCOMS without correlation signal, which indicates good capability of the dynamic range measurement. To confirm the reliability of the FOAC results, the TC of the test pulse is also measured by using a commercial delay-scanning TOAC (Amplitude Technologies Inc., Sequoia-800). Obviously, both the experimental results are in good accordance with each other except for one pre-pulse at -30 ps with approximately $10^{-4}$ intensity, labelled as "a", which only appeared in the measurement profile by using the Sequoia-800. Furthermore, we found a stronger postpulse at the mirror delay time of 30 ps with approximately $0.7 \times 10^{-2}$ intensity, which is labelled as "A". This means that the prepulse "a" is a ghost pulse introduced by postpulse "A" during the TOAC measurement by the Sequoia-800. It should be noted that the ghost prepulse "a" is not detected by the proposed single-shot FOAC, which shows much higher measurement fidelity than that of TOAC. As previously mentioned, this is because the sampling pulse of our single-shot FOAC is generated by a third-order nonlinear process, which has a cleaner TC compared with the sampling pulse generated by SHG in the Sequoia-800 based on TOAC.

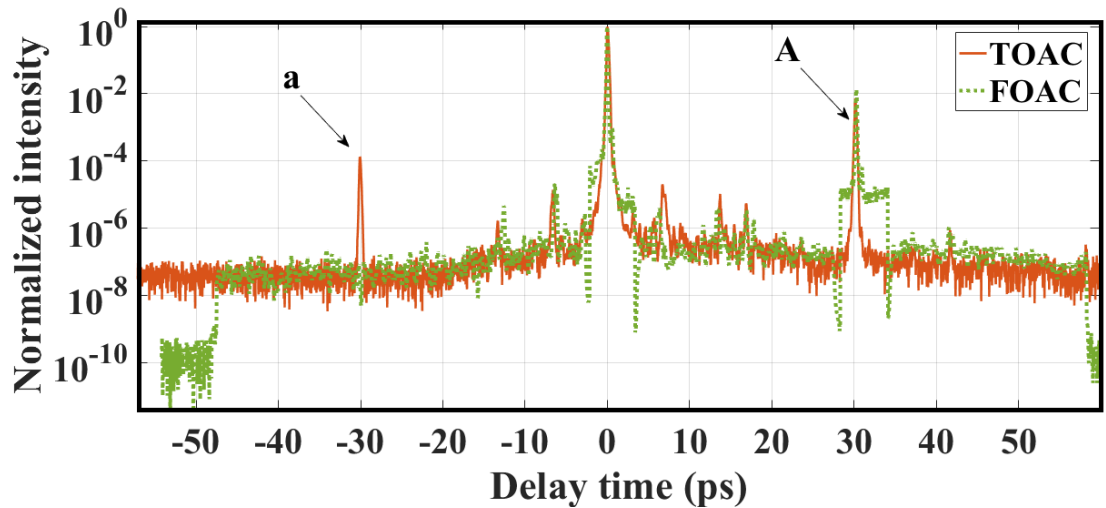

Figure 5. TC results of a Ti: sapphire regenerative amplifier measured using the Sequoia-800 (solid line) and our single-shot FOAC equipment (dash line).

The time resolution is another important parameter in TC measurement. Although the time step can be as short as 10 fs, the time resolution of a delay-scanning TOAC is limited to hundreds femtoseconds by the large GVM between the test pulse and its SHG sampling pulse. In all previous works, the time resolution single-shot TOAC or TOCC measurements was always wider than 500 fs due to GVM and the limitation of detectors. It is because the sampling pulse has the same central wavelength as the test pulse in this single-shot FOAC, the GVM is not a problem. Ideally, a time resolution of approximately 37 fs can be obtained when considering an 11 × 11 μm pixel size. The spatial resolution of the 1:1 4f imaging system, which is used to map the correlation signal from the SFM crystal to the sCMOS sensor, limits the time resolution of this proof-of-principle experiment to approximately 160 fs. By enlarging the main peak pulse in Figure 5 with a linear intensity scale, the correlation profiles by the Sequoia-800 and by single-shot FOAC are shown in Figure 6, where the pulse duration of the test pulse is approximately 40 fs. A high spatial resolution of the 1:1 4f imaging system with a shorter focal length or larger NA are expected to improve the time resolution of this FOAC. A magnified 4f imaging system can also improve the time resolution; however, the time window will be narrowed at the same time. A wider time window can be obtained with a reduced 4f imaging system and a larger camera sensor.

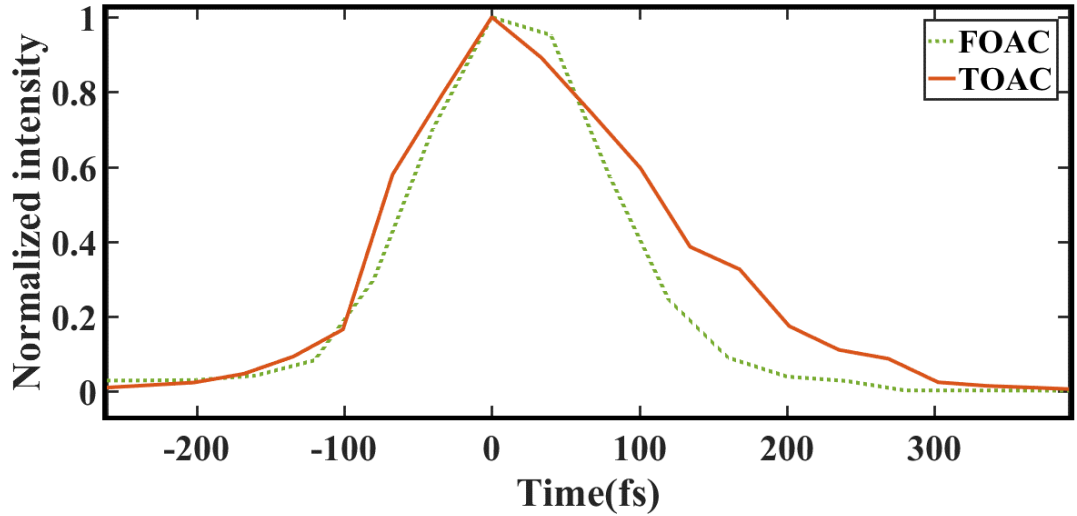

Figure 6. Correlation traces by the TOAC and FOAC with a linear plot of intensity.

To further test the dynamic range capability of our method, we also measured the TC of another laser system using the single-shot FOAC, in which the laser pulse with the parameters of 20 mJ/10 Hz/40 fs/800 nm is used in the experiment. In comparison to the kHz system, both the obtained SD signal and the test pulse have higher pulse energies of 450 µJ and 4 mJ, respectively. As a result, a higher SFM signal can be obtained, which makes it possible to further enhance the dynamic range. The original correlation signal obtained by the sCMOS is shown in Figure 7(a), where two strip-shaped density filters are used to attenuate the main peak signal. A 1-mm glass plate is used to introduce a postpulse for the test pulse, the signal of which is also attenuated by a strip-shaped density filter. The coated wedge is unnecessary in this measurement. Figure 7(b) shows the retrieved TC curves obtained by both the single-shot FOAC and the Sequoia-800. By comparing the intensity of the main peak pulse and the sCMOS noise, a dynamic range of $10^{11}$ is obtained, which, to date, is the highest dynamic range observed for a single-shot TC measurement. In the experiment, to increase the intensity of the correlation signal, the sampling pulse is not expanded by the beam expander to keep its intensity in the overlapping area with the test pulse. As a result, the diameter is not as large as in the above first experiment and the time window is only about 50 ps. Interestingly, the ghost prepulse "a" introduced by the postpulse "A" measured by the Sequoia-800 is not detected by the FOAC again. To be mentioned, the intensity of postpulse "A" introduced by the 1 mm glass plate is measured to be about $0.5 \times 10^{-3}$ at 10.5 ps. The second postpulse at 21 ps is measured to be about $3.2 \times 10^{-7}$ which is in good agreement to the calculated value about $2.5 \times 10^{-7}$. The intensity of the second postpulse is too weak to be observed.

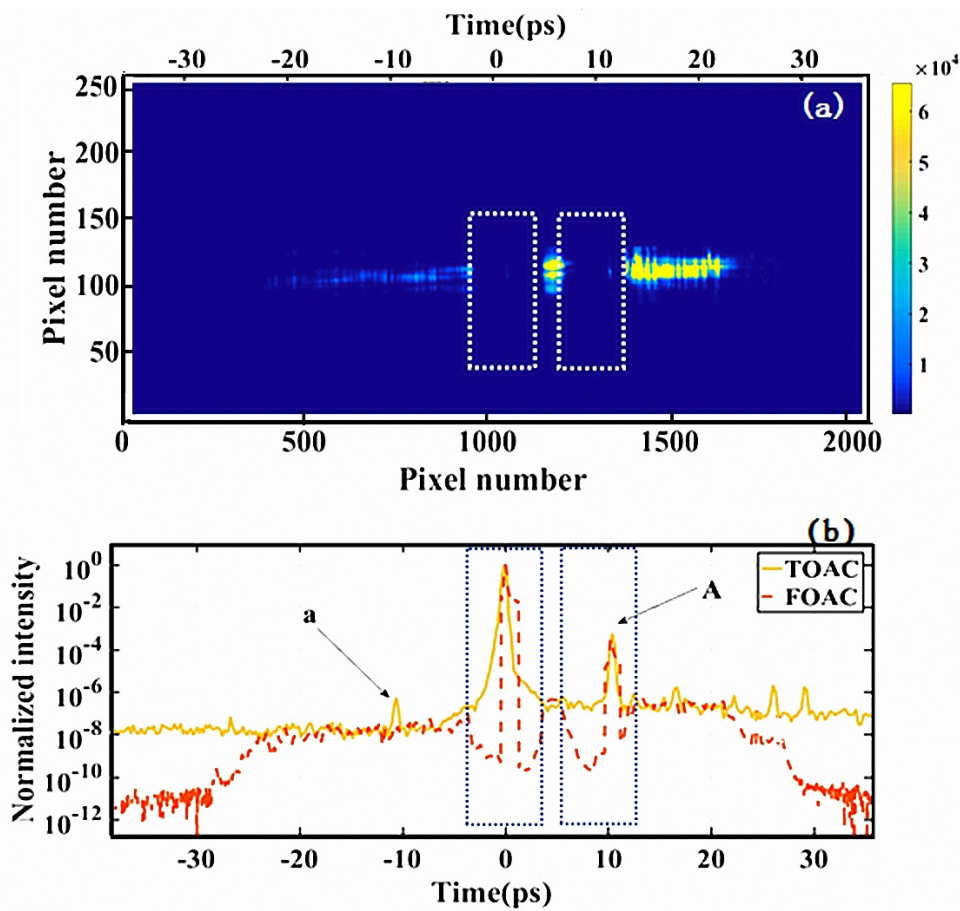

Figure 7. The measurement results with a dynamic range of $10^{11}$ by the FOAC. (a) The correlation signals on the sCMOS detector. (b) The comparison of the TOAC and FOAC measurement results. The strip-shaped density filters are used in the region of the dashed-line rectangle.

As well as for increasing the correlation signal intensity and improving the detector sensitivity, suppressing the scattering noise is also important for a high dynamic range measurement of $10^{11}$. Because the sampling pulse has the same central wavelength as the test pulse in the single-shot FOAC, the correlation signal obtained by SFM has the same central wavelength as the SHG signals of both the sampling and test pulses. Then, the noise introduced by the SHG signals of the sampling and test pulses, which cannot be blocked by a spectral filter, need to be analyzed. It is well known that the SHG process is a second-order optical parametric process that is sensitive to the phase matching condition. Then, a special cutting angle of the BBO is designed to weaken the SHG generation. In the experiment, the crossing angles of the test and sampling pulses with respect to the optic axis of the BBO crystal are 99° and 61°, respectively, which are very different from the best phase matching angle, i.e., 29.2°. This simple design makes the SHG scattering noise in this single-shot FOAC negligible. In the experiment, while blocking either the sampling pulse or test pulse, the intensity of the other SHG scattering noise is captured by the sCMOS camera. The sCMOS background noise is also obtained by covering the shutter of the sCMOS. The results are shown in Figure 8 with different types of lines. The scattering noise from the SHG signals of both incident pulses are at the same scale as the background noise of the sCOMS camera. The results loudly prove the capability of $10^{11}$ dynamic range measurement using our single-shot FOAC system.

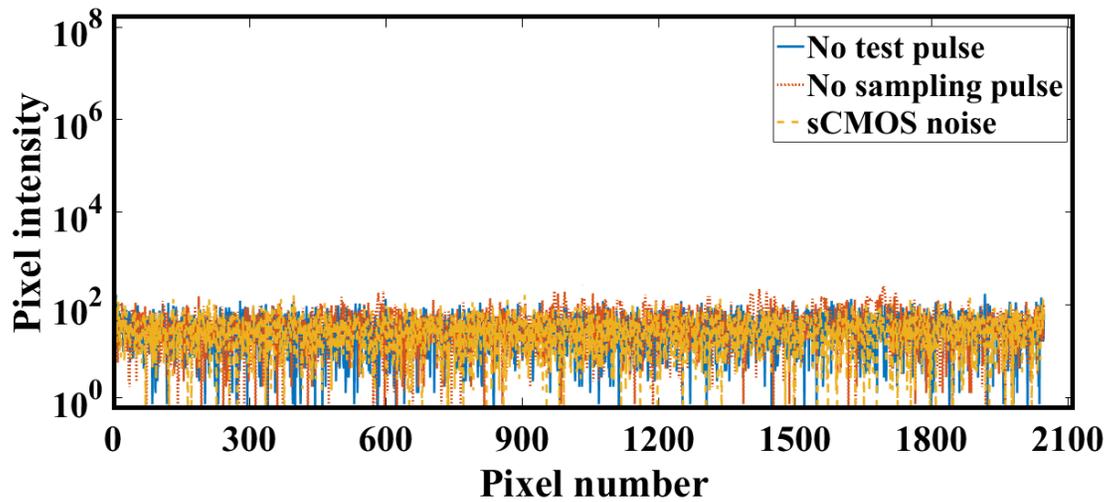

Figure 8. The detector noise from the sCOMS and scattering noise from the SHG signals.

Finally, the single-shot FOAC is tested by using a high TC laser pulse sampled from a PW laser pulse system [32]. The sample pulses have a pulse energy of 10 mJ, repetition rate of 1 Hz, pulse duration of 40 fs, and a central wavelength of 800 nm. Two strip-shaped filters are included to reduce the main peak correlation signal in the experiment. The attenuation ratio of the filter is so large that only the main pulse can be observed and the information near the main pulse is lost, which would miss dynamic ratio data within 5 ps time window around the main pulse. In the future, we would try to solve this problem by fabricating a step variable density filter. Figure 9. (a) shows the original data obtained by the single-shot FOAC. To make the weak signal more visible, the range of the color bar is adjust to be between 100 and 200. And the signal is shown in the Figure 9. (b). The TC curves of the sample pulse measured by the single-shot FOAC and Sequoia-800 are presented in Figure 9.(c). It can be seen that the two measurements in the time window from -40 to 10 ps are consistent with each other very well. A single-shot measurement the TC of a PW laser system with a dynamic range of about $2 \times 10^{10}$ and a time resolution of about 160 fs is obtained, which also shows the capability of $10^{11}$ dynamic range. The correlation signal intensity is dependent on the intensities of the two incident pulses in the nonlinear crystal. Then the intensity profile of the two input beams on the BBO crystal is important for measurement accuracy of the dynamic range. It should be note that the beam profile of the SD signal for sampling pulse after the third order nonlinear process is uniform around the focused plane (the surface of the BBO crystal). Since the test pulse is picked out from the edge of the laser beam, the intensity distribution across the beam will not be uniform. The relatively weaker beam focused from the edge part will induce a weakened SFG signal, this may be the reason why there is large different between the TOAC and FOAC for the data beyond 10 ps. Unfortunately, we did not measure the beam profile of the test beam on the BBO crystal this time. We expect to monitor the beam profiles of the two input beams on the SFG nonlinear crystal by using CCD cameras in the future. Then the intensity distributions of them will be used to correct the final results.

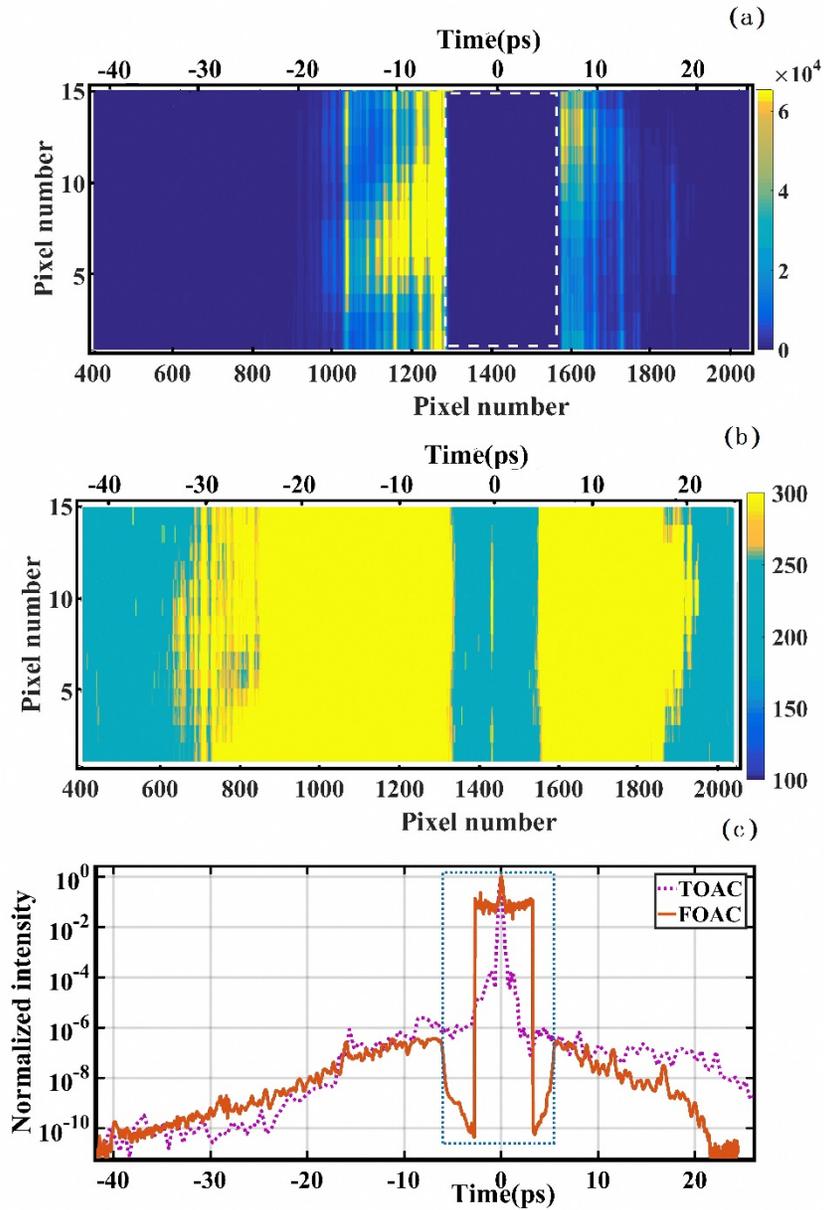

Figure 9. The measurement results of the single-shot FOAC and delay-scanning TOAC (Sequoia-800). (a) The correlation signals on the sCMOS detector. (b). The correlation signals on the sCMOS detector when the range of the color bar is adjusted to make the weak signal more visible. (c). The comparison of the TOAC and FOAC measurement results. The strip-shaped density filters are used in the region of the dashed-line rectangle.

## 6. Conclusion

We have proposed a novel and simple fourth-order autocorrelation method for single-shot TC measurement with higher time resolution and better fidelity than typical third-order autocorrelation or cross-correlation methods. A single-shot FOAC is developed based on a SD process to generate clean sampling pulses. The proof-of-principle experimental results show a dynamic range of approximately $10^{11}$, a time resolution of approximately 160 fs, and a time window of 65 ps capabilities of this single-shot FOAC. Higher dynamic ranges can be obtained in the future by increasing the thickness of the SFM crystal and improving the sensitivity of the detector. A higher time resolution or a wider time window can be achieved by using a magnified or reduced 4f mapping setup in front of the camera. The novel method and the corresponding single-shot FOAC device will benefit the improvement of PW laser systems and many important ultra-high intense laser-matter interaction research activities such as the generation and acceleration of protons and ions, laboratory astronomy, fast-ignition inertial confinement fusion, and the generation of a secondary sources of high-intensity γ-rays.

**Acknowledgments**

This work is supported by the National Natural Science Foundation of China (NSFC) (61527821, 61521093), the Instrument Developing Project (YZ201538) and the Strategic Priority Research Program (XDB160106) of the Chinese Academy of Sciences (CAS), and Shanghai Municipal Science and Technology Major Project (2017SHZDZX02).

The authors would like to thank Dr. Xiaoming Lu, Dr. Xinliang Wang, and Dr. Lingang Zhang for providing the intense laser sources for the single-shot FOAC measurement.